\documentclass[journal]{IEEEtran}
\usepackage{amsmath,amssymb,amsfonts}
\usepackage{algorithmic}
\usepackage{algorithm}
\usepackage{array}
\usepackage{textcomp}
\usepackage{stfloats}
\usepackage{url}
\usepackage{verbatim}
\usepackage{subcaption}
\usepackage{graphicx}
\usepackage{lipsum}
\usepackage{flushend}
\usepackage{booktabs}
\usepackage{enumitem}
\usepackage{epstopdf}

\usepackage[noadjust]{cite}

\usepackage[utf8]{inputenc}
\usepackage[normalem]{ulem}
\usepackage{textcomp}
\usepackage{xcolor}
\usepackage{subcaption}
\usepackage[hidelinks]{hyperref}
\usepackage[font={small}]{caption}

\begin{document}

\title{\huge Received Signal and Channel Parameter Estimation in Molecular Communications}

\author{O. Tansel Baydas,~\IEEEmembership{Student Member,~IEEE,}
        Ozgur~B.~Akan,~\IEEEmembership{Fellow,~IEEE}
\thanks{O.~T.~Baydas and O.~B.~Akan are with the Center for neXt-generation Communications (CXC), Department of Electrical and Electronics Engineering, Ko\c{c} University, Istanbul 34450, Turkey
(e-mail: osman.baydas@boun.edu.tr, akan@ku.edu.tr).}
\thanks{O. B. Akan is also with the Internet of Everything (IoE) Group, Electrical Engineering Division, Department of Engineering, University of Cambridge, Cambridge CB3 0FA, UK (e-mail: oba21@cam.ac.uk).}
\thanks{This work was supported by AXA Research Fund (AXA Chair for Internet of Everything at Ko\c{c} University).}
\thanks{An earlier version of this work was presented at ACM NanoCom’22, Barcelona, Spain \cite{Das2022}.}}

\maketitle

\begin{abstract}

Molecular communication (MC) is a paradigm that employs molecules as information transmitters, hence, requiring unconventional transceivers and detection
techniques for the Internet of Bio-Nano Things (IoBNT). In this study, we provide a novel MC model that incorporates a spherical transmitter and receiver with partial absorption. This model offers a more realistic representation than receiver architectures in literature, e.g. passive or entirely absorbing configurations. An optimization-based technique utilizing particle swarm optimization (PSO) is employed to accurately estimate the cumulative number of molecules received. This technique yields nearly constant correction parameters and demonstrates a significant improvement of 5 times in terms of root mean square error (RMSE). The estimated channel model provides an approximate analytical impulse response; hence, it is used for estimating channel parameters such as distance, diffusion coefficient, or a combination of both. We apply iterative maximum likelihood estimation (MLE) for the parameter estimation, which gives consistent errors compared to the estimated Cramer-Rao Lower Bound (CLRB).
    
\end{abstract}
\begin{IEEEkeywords}
Molecular Communication, Channel Modeling, Received Signal Estimation, Channel Parameter, Maximum Likelihood Estimation
\end{IEEEkeywords}
\maketitle
\section{Introduction}

The combination of nanomachines and biological entities within the Internet of Bio-Nano Things (IoBNT) framework exhibits considerable promise for developing novel applications, including nano biosensors and artificial cells \cite{akan2022internet}, \cite{civas2023graphene}. Molecular communication (MC) is crucial in facilitating those applications \cite{7060516}, \cite{9900784}. The transmission of information in MC is achieved by utilizing molecules, which may be encoded based on type or concentration. 

The use of concentration-based MC is prevalent due to their straightforwardness. However, these methods are susceptible to inter-symbol interference (ISI) caused by the stochastic characteristics of Brownian motion. The presence of interference significantly affects the performance of 
systems, and it is detrimental 
to accurately describe the reception of molecules in order to design the appropriate equipment for an efficient MC channel. Previous research has frequently employed passive receiver (Rx) architectures due to their ease of modelling. Nevertheless, this particular architectural design is physically 

\begin{figure}[htb!]
    \centerline{\includegraphics[scale=1]{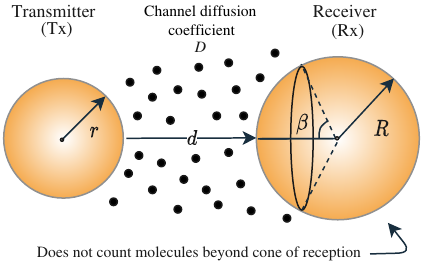}}
    \caption{Molecular SISO channel with spherical Tx and Rx.}
    \label{fig:allmodel}
\end{figure}

separate from the MC channel, hence, limiting its practical importance. In light of this, the active Rx designs are also subject to investigation. In \cite{Akdeniz2018}, the authors proposed the implementation of a spherical receiver which would facilitate the intake of molecules. Machine Learning (ML) approaches were applied as a data-driven method for a 
receiver-transmitter pair with compensating parameters \cite{Yilmaz2017}. The authors also created an artificial neural network (ANN)-based receiver that delivered the same performance as a receiver with a full understanding of the underlying channel model \cite{qian2018receiver}. Furthermore, for a multiple-input multiple-output (MIMO) MC channel with ligand receptors, we proposed the channel modeling and ML-based symbol detection in \cite{baydas23}.

Besides estimating received signals in an MC channel, the recent literature includes the estimation of many channel characteristics, including distance, diffusion coefficient, flow velocity, and the performance restrictions associated with calculated parameters. In \cite{8469051}, authors used a closed-form estimator of the inverse Gaussian channel to estimate the distance between transmitter-receiver pairs. This estimation considered the presence of an unknown diffusion coefficient or drift velocity for the ring-shaped receiver distribution. The authors in \cite{baidoo2020channel} employed the triangular-based technique, which integrates the principles of least squares and gradient descent methods, to estimate the geographical coordinates of the target node (TN). An estimation method is presented in \cite{9767415} that determines the position of a point Tx in a multi-channel (MC) system with numerous fully-absorbing Rxs situated on the surface of spherical arrays. To execute the procedure, the distances between the transmitter and the FA receivers are initially determined and subsequently employed in a multilateration methodology.

Similarly, in \cite{9157845}, the inverse Gaussian channel's closed-form estimator is used to calculate the distance between Tx and Rx.

The Fisher information matrix and the Cramer-Rao lower bound are derived in \cite{9767415} for the joint estimation of multiple channel parameters. This model assumes point transmitters with a non-absorbing receiver, which is unrealistic. In a similar channel, \cite{KUMAR2023} estimates channel parameters using an iterative maximum likelihood estimation (MLE) approach and derive the Cramer-Rao lower bound for the estimation error. The authors assume the probability density function is Gaussian, which also oversimplifies the MC received signal analysis.

In this paper, we model a single-input single-output (SISO) diffusive MC channel, where the receiver is partially absorbing. We provide a novel parameter named 'angle of reception' for the receiver, in addition to the existing correction factors. Particle Swarm Optimisation (PSO) is a rapid and efficient technique to determine optimal values of these parameters. Once it has been demonstrated that the additional parameters for the received signal model remain almost stable across various channel characteristics, the unknown channel parameters, such as distance and diffusion coefficient, are next estimated using Maximum Likelihood Estimation (MLE), a method that has so far been for passive receivers in literature. 


Our contributions are as follows:

\begin{itemize}
    \item We model a realistic SISO MC model with a sphrical transmitter and a partially absorbing spherical receiver.
    \item The proposed PSO-based derivation gives nearly constant correction parameters, which enables us to use this received signal model as an estimator for channels with unknown distance and diffusion coefficient.
    \item We apply an iterative maximum likelihood-based estimation for the channel parameters. We analyze this parameter estimator for three cases: (i) only the distance ($d$) is unknown, (ii) only the diffusion coefficient ($D$) is unknown, and (iii) both $d$ and $D$ are unknown. We compare the normalized error with the Cramer-Rao Lower bound.
\end{itemize}

\section{Channel Model}

We consider a transmitter (Tx) and a receiver (Rx) that are both spherical. These two entities are positioned at a distance of \textit{d} from each other within a medium characterized by a diffusion coefficient of \textit{D}, as shown in \ref{fig:allmodel}. Our approach differs from prior studies, such as those in \cite{Yilmaz2017} and \cite{Farsad2018}, which assume that the receiver is completely absorbing. In our proposed model, the receiver exhibits partial absorption of molecules, limited to a receiving angle denoted as $\beta$. Additionally, it is assumed that the orientation of the transmitter is directly aligned with the receiver, and the respective radii ($r$ and $R$) are within the range of $5$ to $10\mu$m.
In the scenario of perfect conditions when complete absorption occurs, the proportion of molecules that reach the receiver up to a specific time $t$ may be expressed as \cite{Yilmaz2017},

\begin{equation}
    F(t)=\frac{R}{d+R}\times \text{erfc}\left(\frac{d}{y}\right),
\label{eq:1}
\end{equation}
where the function $erfc(\cdot)$ represents the complementary error function, whereas the variable $y$ is defined as the square root of $\sqrt{4Dt}$. Nevertheless, this particular model tends to provide imprecise outcomes, namely, a high root mean square error (RMSE), when dealing with short distances between the transmitter and receiver. To enhance the precision of the model, it is necessary to make adjustments to (\ref{eq:1}) by integrating a cone with an angle denoted as $\beta$. This modification ensures that the receiver's contribution outside the boundaries of this cone may be considered neglected. 
Therefore, in this particular scenario, the cumulative number of molecules received by the receiver may be expressed as \cite{Akdeniz2018},

\begin{equation}
    F(\beta,t)=\frac{R}{d+R}\times erfc\left(\frac{d}{y}\right)\times\frac{\Omega(t,y,\beta)}{U(t,y)},
\label{eq:2}
\end{equation}
where $\Omega$ provides the contribution of molecules observed only in the range of $\beta$, i.e.,
\begin{equation}
\begin{split}
    \Omega(t,y,\beta)=\frac{1}{d}\text{erfc}\left(\frac{d}{y}\right)-\frac{1}{x(\beta)}\text{erfc}\left(\frac{x(\beta)}{y}\right)
    \\
    +\frac{1}{\sqrt{2\pi D t}} \left[\text{Ei}\left(-\frac{d^2}{y^2}\right)-\text{Ei}\left(-\frac{x^2}{y^2}\right)\right],
\end{split}
\end{equation}
where $x(\beta)$ is the distance between the transmission point and the boundary of the receiving cone, i.e., 
\begin{equation*}
    x(\beta)=\sqrt{(R+d)^2-2(R+d)R\cos{(\beta)}+R^2},
\end{equation*}
and
\begin{equation}
    \begin{split}
    U(t,y)=\frac{1}{d}\text{erfc}\left(\frac{d}{y}\right)-\frac{1}{d+2R}\text{erfc}\left(\frac{d+2R}{y}\right)
    \\
    +\frac{1}{\sqrt{D t}} \left[\text{Ei}\left(-\frac{d^2}{y^2}\right)-\text{Ei}\left(-\frac{(d+2R)^2}{y^2}\right)\right],
\end{split}
\end{equation}
in which $Ei(\cdot)$ is the exponential integral function.

We modify the channel model in (\ref{eq:2}) to make it more realistic by adding three correction parameters: a global scaling factor $b_1$ (also known as model fitting), and $b_2$ and $b_3$ that are the exponents of $4D$ and $t$ in $y$, respectively, where $t$ is time. Therefore, we can rewrite (\ref{eq:2}) with the new $b_i$ parameters as follows:

\begin{equation}
    F'(\beta,b_i,t)=b_1\times\frac{R}{d+R}\times \text{erfc}\left(\frac{d}{y_n}\right)\times\frac{\Omega(t,y_n,\beta)}{U(t,y'_n)},
\label{eq:5}
\end{equation}
where $y_n=(4D)^{b_2} t^{b_3}$. Ultimately, our model has a total of four parameters, which are $\beta$, $b_1$, $b_2$, and $b_3$.

The dataset used in this work is obtained from \cite{Yilmaz2017}. This dataset provides information on the number of molecules received by the receiver in a (SISO) channel with a spherical transmitter-receiver pair.
In contrast to the findings presented in \cite{Yilmaz2017}, our study introduces an extra parameter ($\beta$) that is influenced by variables $d$, $R$, and $D$. This poses a challenge in training the parameters using the complete dataset, as this procedure necessitates the independence of all parameters from input variables. Therefore, the model is trained for several scenarios involving varying values of $d$, $R$, and $D$. The mean squared error (MSE) is computed for our proposed model through an exploration of the parameter space, aiming to identify the optimal fit based on the loss function outlined in equation (\ref{eq:LOSS}), i.e.,

\begin{equation}
    \sum_{k=1}^M\left(F'(\beta,b_i,t_k)-S(t_k)\right)^2,
\label{eq:LOSS}
\end{equation}

A total of 200 iterations were conducted for each example of the Particle Swarm Optimiation (PSO) algorithm \cite{PSO}, employing the PSO hyperparameters and dataset parameters as outlined in Table~\ref{Table: Parameters simu}. The root mean square error (RMSE) may be computed by taking the square root of the mean squared error (MSE) calculated using (\ref{eq:LOSS}), which is determined using the best parameters acquired from the particle swarm optimization (PSO) search.

\begin{table}[t!]
\centering
\renewcommand*{\arraystretch}{1.1}
\caption{Parameter values used in simulations.}
\begin{tabular}{ |l|l|l| }
 \hline
 \textbf{Parameter} & \textbf{Value} & \textbf{Unit} \\
 \hline
 \hline
 \textit{Num. of transmitted molecules} & $3000$ & -\\
 \textit{Num. of Rxs} & $5$ & -\\
 \textit{Simulation time step, $\Delta T$} & $50$ & [$\mu$s] \\
 \textit{Tx-to-Rx distance, $d$} & $2$, $4$, $6$, $8$, $10$ & [$\mu$m] \\
 \textit{Rx radii, $R$} & $3$, $5$, $7$ & [$\mu$m] \\
 \textit{Tx radii, $r$} & $3$, $5$, $7$ & [$\mu$m] \\
 \textit{Diffusion coefficient, $D$} & $50$, $100$ & [$\mu$m$^2$/s] \\
 \textit{PSO cognitive parameter} & $0.5$& -\\
 \textit{PSO social parameter} & $0.3$& -\\
 \textit{PSO inertia parameter} & $0.9$& -\\
 \hline
\end{tabular}
\label{Table: Parameters simu}
\end{table}

\begin{figure}[t!]
    \centerline{\includegraphics[scale=0.5]{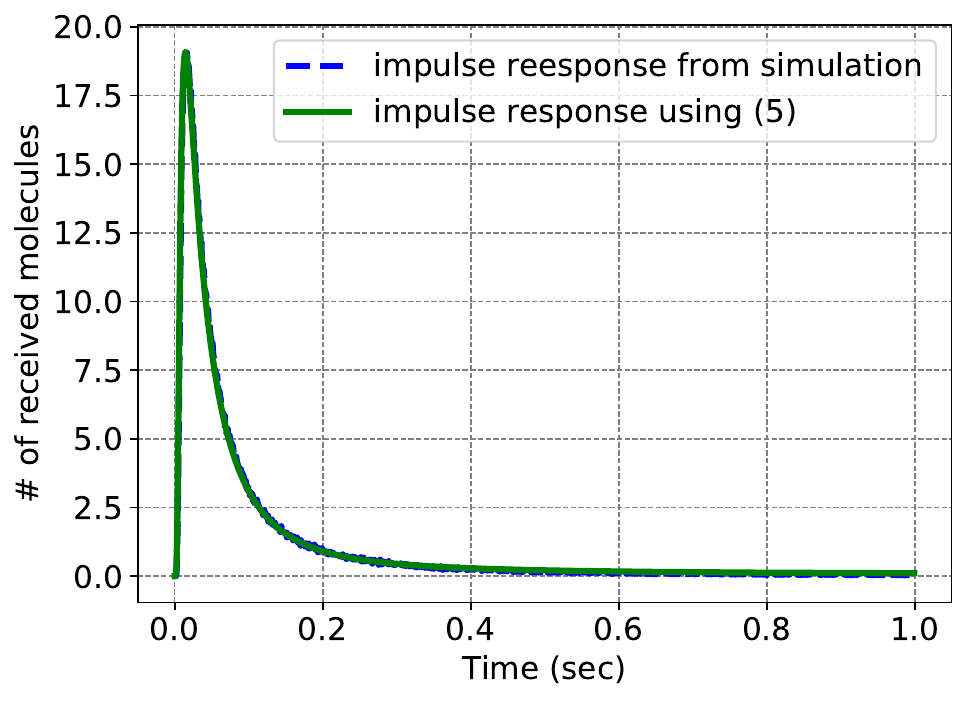}}
   \caption{Impulse response of simulation data and our model.}
    \label{fig:resp1}
\end{figure}

The experimental results demonstrate that the values of $b_1$, $b_2$, and $b_3$ remain relatively consistent across various input parameters, namely $d$, $R$, and $D$. Furthermore, it is evident from the data shown in Table~\ref{Table: RMSE} that the error values are about $0.44$ for nearly all scenarios. This finding supports the claim that the estimated number of received molecules remains consistent across different values of $d$ and $R$, indicating that the distance between the transmitter and receiver or the radius of the receiver does not influence it. This improvement is significant in comparison to the findings of \cite{Yilmaz2017} since their root mean square error (RMSE) is influenced by both the parameter $d$ and the variable $R$, resulting in elevated values, particularly when $d$ is less than $4\mu$m. Therefore, our model demonstrated a significantly reduced root mean square error (RMSE) in estimating the cumulative number of received molecules, with a five-fold decrease compared to the findings reported by \cite{Yilmaz2017}. The alignment between the impulse response of our model and the actual simulation data for a displacement value of $d=2\mu m$ is illustrated in Fig. \ref{fig:resp1}. This comparison demonstrates the reliability and accuracy of our model.

 \begin{table}[t!]
\centering
\caption{RMSE of the estimated number of received molecules using (5).}
 \begin{tabular}{ |l|l|l|l|  }
 \hline
 $\textbf{d/R}$ & \boldmath{${5\mu m}$} & \boldmath{${7.5\mu m}$} & \boldmath{${10\mu m}$} \\
 \hline
 \hline
 \boldmath{${2\mu m}$} & 0.439 & 0.439 & 0.440\\
 \boldmath{${4\mu m}$} & 0.470 & 0.440 & 0.467 \\
 \boldmath{${6\mu m}$} & 0.439 & 0.439 & 0.439\\
 \boldmath{${8\mu m}$}  & 0.440 & 0.438 & 0.470\\
 \boldmath{${10\mu m}$} & 0.440 & 0.439 & 0.440\\

\hline
\end{tabular}
\label{Table: RMSE}
\end{table}

We use our received signal model for the estimation of channel parameters. For the calculation of the probability density function (PDF), the impulse response is needed. To derive the impulse response from the cumulative received molecules with optimized compensation parameters, which is given in (\ref{eq:5}), we approximate the complex functions into analytical functions, which is reasonable to take the derivative of it. Based on \cite{abramowitz1948handbook}, for the $erfc(\cdot)$, we use the approximation as,

\begin{equation}
    erfc(x) \approx (\textbf{a}^{T}\textbf{x}_6)^{-16}  = h_1(x)
\label{eq:7}
\end{equation}
where $\textbf{a} = [a_0, a_1, . . . a_5]$ and $\textbf{x}_6 = [1, x, . . . x^6]$. The coefficients for $\textbf{a}$ are given in Table \ref{Table: erfc}.

\begin{table}[b!]
\centering
\caption{ Coefficients for $erfc(.)$ approximation.}

 \begin{tabular}{ |l|l|l|l|  }
 \hline
 \boldmath{${a_i}$} & \textbf{value}  \\
 \hline
 \hline
 $a_0$ & 0.0705230784 \\
 $a_1$ & 0.0422820123 \\
 $a_2$ & 0.0092705272 \\
 $a_3$  & 0.0001520143 \\
 $a_4$ & 0.0002765672  \\
 $a_5$ & 0.0000430638  \\

\hline
\end{tabular}
\label{Table: erfc}
\end{table}



\begin{figure}[t!]
    \centerline{\includegraphics[scale=0.5]{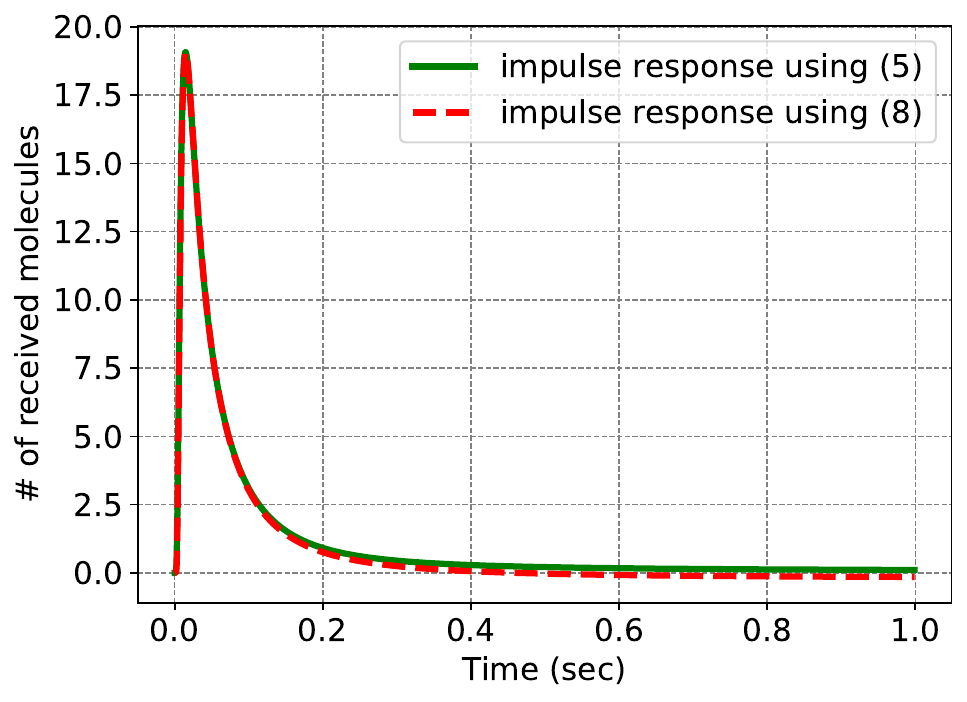}}
   \caption{Comparison of impulse responses in (5) and (8).}
    \label{fig:resp2}
\end{figure}

Considering this approximation, the channel impulse response $f(t)$ is obtained as,

\begin{equation}
\begin{split}
    f(t)=b_1 \times \frac{R}{d+R}\times \left[ -h_{1}'\left(\frac{d}{y_n}\right)\times\frac{dy_n'}{y^2}\times\frac{\Omega(t,y_n,\beta)}{U(t,y_n)} \right. \\ + h_{1}\left(\frac{d}{y_n}\right)\times\frac{\Omega'(t,y_n,\beta)U(t,y_n)- \Omega(t,y_n,\beta)U'(t,y_n)}{U^2(t,y_n)} \Bigg.\Bigg]
\end{split}
\label{eq:8}
\end{equation}
where $y_n' = b_3(4D)^{b_2} t^{b_3-1}$ is the derivative of $y_n$ and $h_1^{'}(\cdot)$ is the derivative of $h_1(\cdot)$. The impulse response is shown in Fig.\ref{fig:resp2}. It can be seen that the approximation does not affect the performance of our model significantly. The RMSE between analytical impulse response given in (\ref{eq:8}) and numeric impulse response calculated from (\ref{eq:5}) is given in Table \ref{Table: RMSE2}.

\begin{table}[b!]
\centering
\caption{ RMSE of the approximated estimated number of received molecules using (\ref{eq:8}).}
 \begin{tabular}{ |l|l|l|l|  }
 \hline
 \textbf{d/D} & \boldmath{${5\mu m^2}$} & \boldmath{${7.5\mu m^2}$} & \boldmath{${10\mu m^2}$} \\
 \hline
 \hline
 \boldmath{${2\mu m}$} & 0.2180 & 0.2598 & 0.3423\\
 \boldmath{${4\mu m}$} & 0.0890 & 0.1424 & 0.1604 \\
 \boldmath{${6\mu m}$} & 0.0149 & 0.03720 & 0.0532\\
 \boldmath{${8\mu m}$} & 0.0050 & 0.0230 & 0.0182\\
 \boldmath{${10\mu m}$} & 0.0018 & 0.0079 & 0.0059\\

\hline
\end{tabular}
\label{Table: RMSE2}
\end{table}

\section{Channel Parameter Estimation}

The receiver's sampling process possesses a sufficient number of timeslots, allowing for the assumption of independence among the observations.  Additionally, it is assumed that the individual observations, which follow a Binomial distribution, may be modelled as Poisson random variables. The means of these Poisson random variables correspond to the expected values of the observations at their respective times. Hence, by utilizing the impulse response provided in (6), it is possible to express the joint probability density function (PDF) of the receiver's observations as follows \cite{Schoberest},
\begin{equation}
    p(\mathbf{s} \mid \boldsymbol{\theta})=\prod_{m=1}^M f\left(t_m\right)^{s_m} \exp \left(-f(t_m)\right) / s_{m} !
\end{equation}
where $\boldsymbol{\theta}= \{d,D\}$, $M$ is the number of samples for one transmission, and $s_m$ is the observation, which is the number of received molecules in $m_{th}$ timeslot.  Let us represent $p(d,D)$ as a log-likelihood function,
\begin{equation}
    \begin{split}
        P(\mathbf{s} \mid \boldsymbol{\theta})=\ln p(\mathbf{s} \mid \boldsymbol{\theta})
        \\
        =\sum_{m=1}^M\left[s_m \ln f\left(t_m\right)
        -\ln s_{m} !-f\left(t_m\right)\right]
    \end{split}
\end{equation}

The partial derivative of the log-likelihood function becomes

\begin{equation}
    \frac{\partial P(\mathbf{s} \mid \boldsymbol{\theta})}{\partial \theta} = \sum_{m=1}^M\left[\frac{s_m}{f\left(t_m\right)}-1\right]f_{\theta}\left(t_m\right)
\end{equation}
where $f_{\theta}\left(t_m\right)$ is the partial derivative of $f\left(t_m\right)$ with respect to $d$ or $D$. Similarly,

\begin{equation}
\begin{split}
       \frac{\partial^2 P(\mathbf{s} \mid \boldsymbol{\theta})}{\partial \theta^2} = \sum_{m=1}^M\left[-\frac{s_m}{f\left(t_m\right)^2}\right]f_{\theta}\left(t_m\right) \\
       + \left[\frac{s_m}{f\left(t_m\right)}-1 \right]f_{\theta \theta}\left(t_m\right)
\end{split}
\end{equation}
where $f_{\theta \theta}$ is the second derivative of impulse response calculated in (8).

We can maximize the likelihood function to estimate

\begin{equation}
   \widehat{\boldsymbol{\theta}}=\{\widehat{d}, \widehat{D}\}=\underset{d, D}{\operatorname{argmax}} P\left(\mathbf{s} \mid d, D\right)
\end{equation}

We use the Newton-Raphson Method for the likelihood maximization. It should be noted that $t_m$ differs for each total sampling number. Furthermore, the first sampling is made when $t_1 = 20 ms$, and the remaining samples are distributed equally until the $t_{end}$. The iteration is stopped when the estimation becomes constant.  

\subsection{Unknown d}
The Newton-Raphson method begins with an initial estimate
$d_0$. The mean of our range of $d$'s is taken as the initial point. For this case, $D$ is known. Then, the $(n+1)^{th}$ estimate is found iteratively as

\begin{equation}
    \hat{d}_{n+1}=\hat{d}_n-\frac{\partial P(\mathbf{s} \mid \boldsymbol{\theta})}{\partial d} /\left.\frac{\partial^2 P(\mathbf{s} \mid \boldsymbol{\theta})}{\partial d^2}\right|_{d=\hat{d}_n}
\end{equation}

\begin{figure*}[t!]
    \subfloat[\centering]{\includegraphics[height=0.17\textheight]{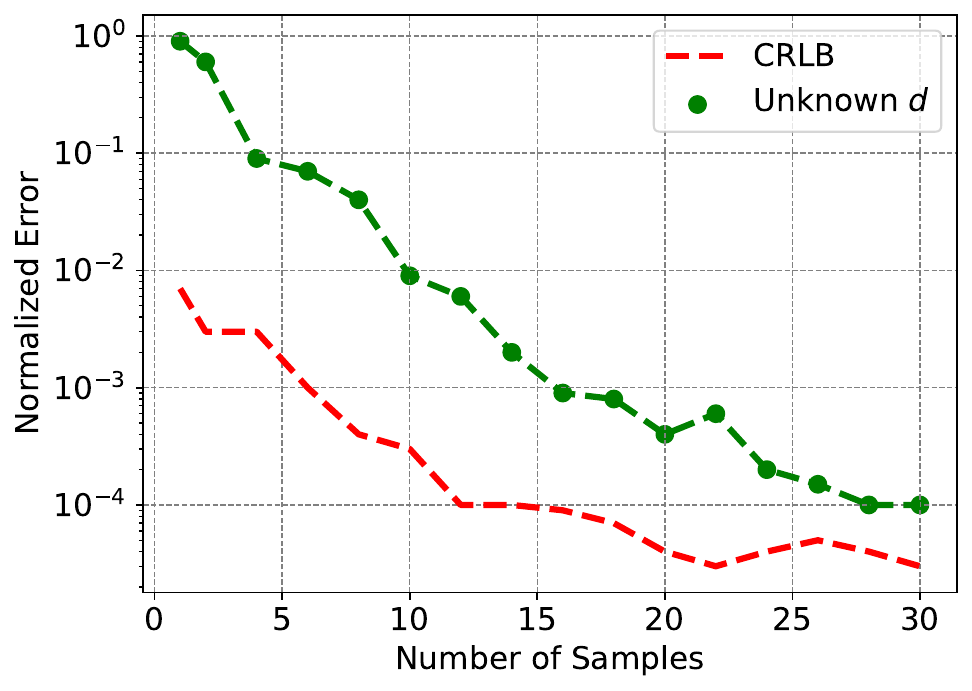}}
    \subfloat[\centering]{\includegraphics[height=0.17\textheight]{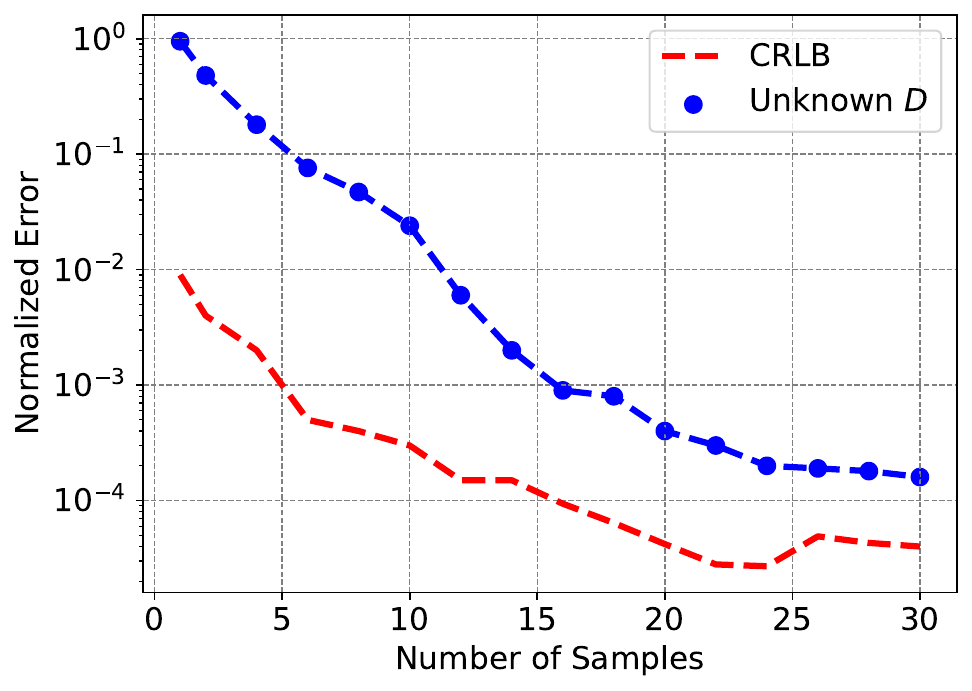}}
    \subfloat[\centering]{\includegraphics[height=0.17\textheight]{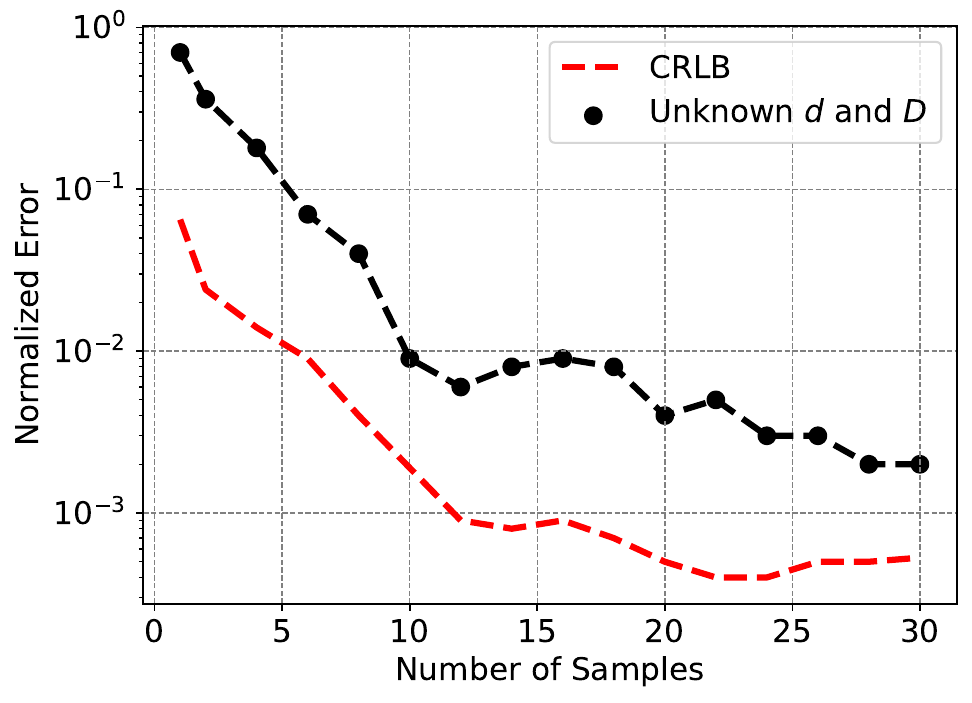}}
    \caption{Normalized error performance of the Nexton-Raphson method for different number of samples and different cases: a) unknown $d$, b) unknown $D$, and c) unknown $d, D$.}
    \label{fig:newtonerror}
\end{figure*}

\subsection{Unknown D}

We assume that $d$ is known. Similar to the previous case, $(n+1)^{th}$ estimate is found iteratively as

\begin{equation}
    \hat{D}_{n+1}=\hat{D}_n-\frac{\partial P(\mathbf{s} \mid \boldsymbol{\theta})}{\partial D} /\left.\frac{\partial^2 P(\mathbf{s} \mid \boldsymbol{\theta})}{\partial D^2}\right|_{D=\hat{D}_n}
\end{equation}
\subsection{Unknown d and D}

When both distance and the diffusion coefficient are unknown, we use the joint estimation, which can be expressed as
\begin{equation}
\begin{split}
\left[\begin{array}{l}
\hat{d}_{n+1} \\
\hat{D}_{n+1}
\end{array}\right]=\left[\begin{array}{l}
\hat{d}_n \\
\hat{D}_{n}
\end{array}\right]-\left[\begin{array}{ll}
\frac{\partial^2 P(\mathbf{s} \mid \theta)}{\partial d^2} & \frac{\partial^2 P(\mathbf{s} \mid \theta)}{\partial d \partial D} \\
\frac{\partial^2 P(\mathbf{s} \mid \boldsymbol{\theta})}{\partial d \partial D} & \frac{\partial^2 P(\mathbf{s} \mid \theta)}{\partial D^2}
\end{array}\right]^{-1}\\
\times \left[\begin{array}{l}
\frac{\partial P(\mathbf{s} \mid \theta)}{\partial d} \\
\frac{\partial P(\mathbf{s} \mid \theta)}{\partial D}
\end{array}\right]
\end{split}
\end{equation}

\subsection{Cramer-Rao Lower Bound}

The Fisher information matrix is given by
\begin{equation}
    \mathbf{I}(\theta)_{u w}=-\mathbb{E}\left[\frac{\partial^2  P\left(\mathbf{s} \mid \theta\right)}{\partial \theta_u \theta_w}\right],
\end{equation}
where $\theta$ is the parameters to be estimated. The expectation is calculated with respect to $\textbf{s}$.

For the unknown $d$ or $D$ case, the Fisher information becomes
\begin{equation}
\mathbf{I}(\theta) = -\mathbb{E}\left[
\frac{\partial^2 P(\mathbf{s} \mid \theta)}{\partial \theta^2}\right],
\end{equation}
where \textbf{$\theta$} is $d$ or $D$. For the joint estimation of $d$ and $D$, the Fisher information becomes
\begin{equation}
\mathbf{I}(\theta) = -\mathbb{E}\left[\begin{array}{ll}
\frac{\partial^2 P(\mathbf{s} \mid \theta)}{\partial d^2} & \frac{\partial^2 P(\mathbf{s} \mid \theta)}{\partial d \partial D} \\
\frac{\partial^2 P(\mathbf{s} \mid \boldsymbol{\theta})}{\partial d \partial D} & \frac{\partial^2 P(\mathbf{s} \mid \theta)}{\partial D^2}
\end{array}\right].
\end{equation}

The covariance matrix of any unbiased estimator is given as
\begin{equation}
    \mathbf{C}_\theta \geq \mathbf{I}^{-1}(\theta).
\end{equation}

Therefore, the variance, $Vr((\theta)$, is lower bounded by
\begin{equation}
Vr\left(\theta\right)=\left[\mathbf{C}_\theta\right]_{u w} \geq\left[\mathbf{I}^{-1}(\theta)\right]_{u w}.
\end{equation}

The Normalized Mean Square Error (NMSE) is calculated from (13), (14), and (15). The results are compared with the normalized CRLB given in (20) and shown in Fig. \ref{fig:newtonerror}. Our estimation is consistent with the Cramer-Rao Lower bound, giving approximately $10^{-4}$ error in $~15$ samples. The error of joint estimation is higher than the cases only $d$ or $D$ are estimated, which is expected from CRLB. 

\section{Conclusions}

This work presents a PSO-based channel impulse derivation method that yields almost constant correction parameters. Furthermore, a 5 times improvement in the root mean square error (RMSE) of the estimated number of received molecules was attained, regardless of the distance and radius of the receiver (Rx) and transmitter (Tx). 
Using our approximated channel impulse response model and employing the iterative Maximum Likelihood Estimation (MLE) technique, we have successfully derived estimations for both distance and diffusion coefficient. Notably, our estimations have consistent error performance when compared to the computed Cramer-Rao Lower Bound (CLRB) and passive receiver models in the literature. This study extends the existing research by introducing the application of the maximum likelihood estimate (MLE) approach to actively absorbing receiver structures, which are considered to be more physically realistic than passive receivers.  As further work, we plan to investigate this model as a closed-form method instead of iterative MLE and applications of our model to ligand receptor-based receivers.

\bibliographystyle{IEEEtran}
\bibliography{References}

\end{document}